# Enhancement of the Lifetime of Metastable States in Er-doped Si Nanocrystals by External Colored Noise

**S. Spezia*[1], D. Valenti[1], D. Persano Adorno[1] and B. Spagnolo[1,2]**

[1]Dipartimento di Fisica e Chimica, Group of Interdisciplinary Theoretical Physics
Università di Palermo and CNISM, Unità di Palermo,
Viale delle Scienze, Edificio 18, I-90128 Palermo, Italy

[2] Istituto Nazionale di Fisica Nucleare, Sezione di Catania,
Via S. Sofia 64, I-90123 Catania, Italy

## ABSTRACT

The changes in the lifetime of a metastable energy level in Er-doped Si nanocrystals in the presence of an external source of colored noise are analyzed for different values of noise intensity and correlation time. Exciton dynamics is simulated by a set of phenomenological rate equations which take into account all the possible phenomena inherent to the energy states of Si nanocrystals and $Er^{3+}$ ions in the host material of Si oxide. The electronic deexcitation is studied by examining the decay of the initial population of the Er atoms in the first excitation level $^4I_{13/2}$ through the fluorescence and the cooperative upconversion by energy transfer. Our results show that the deexcitation process of the level $^4I_{13/2}$ is slowed down within wide ranges of noise intensity and correlation time. Moreover, a nonmonotonic behavior of the lifetime with the amplitude of the fluctuations is found, characterized by a maximum variation for values of the noise correlation time comparable to the deexcitation time. The indirect influence of the colored noise on the efficiency of the energy transfer upconversion activated from the level $^4I_{13/2}$ is also discussed.

---

* Corresponding Author address. Email: stefano.spezia@gmail.com.

# INTRODUCTION

The maximum efficiency of the photovoltaic cells, consisting of single-junctions, is limited to about 30% (for one sun illumination) by intrinsic losses such as the inability to absorb below-band gap photons. This intrinsic characteristic, called the Shockley-Queisser limit, is due to the discrete nature of the band structure of semiconductors. [1-7] Only photons of energy at least equal to the band gap are absorbed and can contribute to the photogeneration of electrons in a photovoltaic device. The photons with energy lower than the energy band gap are transmitted through the solar cell and do not contribute to the output electric power. [5]

A way to exceed this limit is to utilize the low energy photons via electronic *upconversion* process. [3, 5-6] Upconversion processes of photons of low energy into photons of higher energy offer a promising possibility of increasing the efficiency of a solar cell, allowing to exploit also photons of the solar spectrum with energy lower than the band gap of the material used. In particular, detailed balance calculations have shown that the application of an *upconverter* to a solar cell increases its efficiency limit from 31% to 37.4%. [3] Therefore, a full investigation of the efficiency of an upconverting material becomes an essential point in the design of a solar cell/upconverter integrated system.

An upconverter consists of a material with a band gap almost equal to that of the solar cell and containing intermediate levels. The upconverters usually combine an active ion, whose energy level scheme is employed for absorption of low energy photons, and a host material, in which the active ion is embedded. The most efficient upconversion has been found by using rare earths, i.e. lanthanides elements such as Erbium, Ytterbium, etc. [7-8] Moreover, the silicon nanocrystals (Si nc) in silicon oxide represent a possible candidate to incorporate rare earths, such as erbium atoms (Er), due to their compatibility with the silicon technology.

Among the upconversion mechanisms, a crucial role is played by the energy transfer upconversion (ETU) processes, which are based on cooperative effects. ETU processes have demonstrated the highest efficiency with respect to other mechanisms, i.e. two-photon absorption, which request very high power of the incident radiation. Moreover, Pilla et al. and Jacinto et al. have recently showed that the quantum efficiency of the fluorescence of a given electronic energy level is directly related to both the lifetime of the considered metastable state and the rate of ETU process. [9-10]

Over the last few decades experimental and theoretical studies have more and more stressed the importance of the noise fluctuations in natural and artificial systems. Noise-induced effects have been experimentally observed and theoretically studied in different physical and biological contexts. [11-28] In particular, stochastic resonance, [24-28] resonant activation [29-31] and noise enhanced stability [32-38] phenomena in several systems have been widely discussed. Specifically, the noise enhanced stability (NES) effect implies that a system remains in a metastable state for a time longer than the deterministic counterpart, and the lifetime has a maximum in correspondence with a well determined noise intensity. [34-35, 38]

In a previous work, Pacifici et al. have discussed a possible phenomenological description of the Er-doped Si nanocrystals, based on a set of rate equations which allow to quantitatively justify the results of optical measurements. [39]

The purpose of the present work is to study the effects of an external colored noise on the deexcitation dynamics of the first excited energy level of $Er^{3+}$ ions embedded in Si nc. By superimposing an external noise source to the intrinsic one, it is possible to modify the dynamic response of the system. The intrinsic noise, which in principle could be modeled as a white noise, is already taken into account in the transition rates and represents the effect of the host temperature. In particular, by using a stochastic model based on rate equations, we focus on the influence of a multiplicative colored noise, modeled as an Ornstein-Uhlenbeck process. [40] As we have discussed, the physical condition of maximizing the efficiency of the upconversion phenomenon coincides with that of enhancing the average lifetime of particular energy levels of the Er atoms. The metastable state of interest is the first excited level of $Er^{3+}$ ions, on which the effects of the external colored noise have been investigated by solving numerically the equations of the model. Our findings show that the presence of the external source of colored noise can determine an increase of the deexcitation time. In particular, this increase is maximum at specific values of the noise intensity and dependent on the ratio between the noise correlation time and the characteristic time of the deexcitation process obtained in the absence of fluctuations.

# Er-DOPED Si NANOCRYSTALS

## Si Nanocrystal-Er Atom Interaction

In 2003, photoluminescence measurements have allowed to identify a schematic diagram of the energy levels for the system of Si nanocrystals doped with Er atoms [11] (see Figure 1).

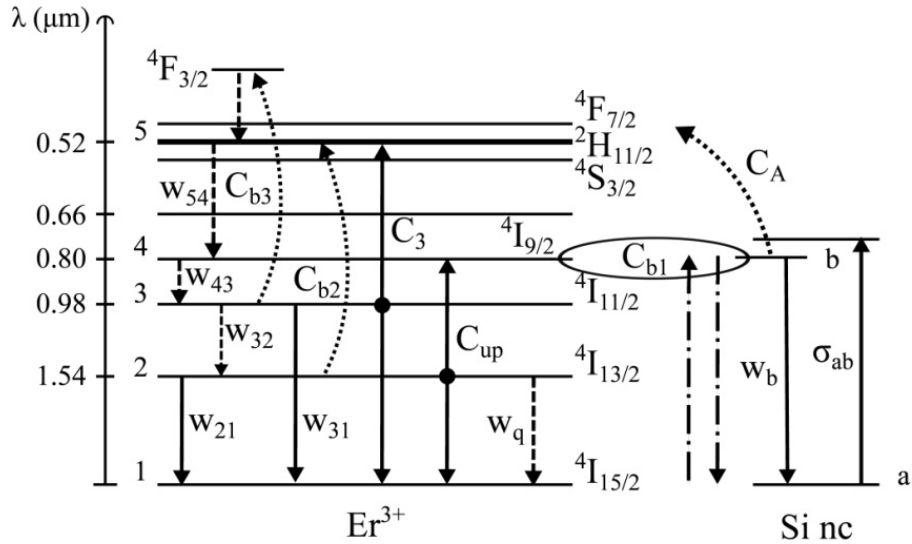

Figure 1. Energy level scheme for the Er-doped Si nanocrystals system. The Si nanocrystals are modeled by a two-effective energy level scheme, while $Er^{3+}$ ions are represented as a five-effective energy level diagram.

In this scheme, a Si nanocrystal (nc) is represented by a three level system, consisting of two band edge levels and an interfacial level, which acts as a trap for the exciton. [41–43] Since we assume the trapping phenomenon to be a very fast process in comparison with the typical transition times, we can consider the Si nc as an effective two-level system, where the ground state is represented by the level *a* and the upper energy state by the level *b*. Within this assumption, $\sigma_{ab}$ is the effective cross section of excitation, and describes both the creation of an exciton and its subsequent trapping at the interfacial energy level. Both processes occur after the absorption of a 488 nm photon. $w_b$ is the total recombination rate of an exciton for an isolated Si nc, and is given by the sum of both radiative and non-radiative recombination rates. The $Er^{3+}$ ion is described as a five-effective level system. In fact, due to the Stark splitting, the fifth level is composed of $^4S_{3/2}$, $^2H_{11/2}$, and $^4F_{7/2}$, which are strongly overlapping, and can be therefore taken into account as a single level. $w_{ij}$ indicate the total transition rates from level *i* to level *j*, where $i, j = 1, 2, ..., 5$ and $i > j$.

All the parameters of Figure 1 are listed in Table 1. Specifically, we can distinguish two classes of parameters: the transition rates expressed in $s^{-1}$ and the coefficients describing energy transfers expressed in $cm^3 s^{-1}$. Among the second-class parameters, the coefficient $C_{b1}$ describes the coupling between the excited level of Si nc and the Er ground state, which is responsible for the energy transfer between the Si nc's and Er ions surrounding it. This energy transfer is represented by the two antiparallel arrows, standing for the non-radiative deexcitation of an excited Si-nc (down arrow) and the following $Er^{3+}$ excitation to the $^4I_{9/2}$ level (up arrow). The parameters $C_{b2}$ and $C_{b3}$ describe the excitation of level 5 of $Er^{3+}$ from the levels *2* and *3,* respectively. In other words, these coefficients describe the further excitations of an excited $Er^{3+}$ state through the energy transfer from the excited level *b* of the Si nc. The constant $C_A$ takes into account a phenomenon very similar to the Auger effect occurring in Er-doped crystalline Si (Ref. [44]). In the Er-Si nc system this effect consists in an energy transfer from an excited Er level back to a confined exciton of a Si nc, which is therefore promoted to a higher energy level. This effect is present only at high pump powers and mainly involves the long lived $^4I_{13/2}$ level. The scheme also represents the quenching effect, which is due to the energy migration over the sample due to the energy transfer between two neighbouring $Er^{3+}$ ions, one in the first excited level, the other in the ground state. The value of the total rate of the deexcitation from the first excited Er level is $w_{21} + w_q$, with $w_{21} = 4.2 \times 10^2$ s$^{-1}$. The rate $w_q = 8\pi C_{Er} N_q N_0$ takes into account the quenching effect, where $N_q$ is the concentration of the quenching centers, $C_{Er}$ is the coupling coefficient describing the interaction between the ground state and the first excited level of Er, and $N_0$ is the total Er concentration.

The value of the product $C_{Er} N_q$ has been obtained by experiments[39] and it is equal to $3.2 \times 10^{-20}$ cm$^3$ s$^{-1}$. Other two processes are represented by a double arrow (with a single point of origin), and are indicated by the constants $C_{up}$ and $C_3$. They are the cooperative up-conversion coefficients describing the interaction of two neighbouring $Er^{3+}$ ions which are both in the first or in the second excited state, respectively. In the first case one of the two ions decays to the ground state giving its energy to the other one, which will be excited to the $^4I_{9/2}$ level. In the second case the interaction will bring an ion to the $^2H_{11/2}$ level and the other one to the ground state.

**Table 1. Parameters of the physical model**

| Symbol | Value | Reference |
|---|---|---|
| $\lambda_{pump}$ | 488 nm | Ref. [39] |
| $\sigma_{ab}$ | $2 \times 10^{-16}$ cm$^2$ | Ref. [45] |
| $\sigma$ | $1 \times 10^{-19}$ cm$^2$ | Ref. [45] |
| $w_b$ | $2 \times 10^4$ s$^{-1}$ | Ref. [45] |
| $w_{21}$ | $4.2 \times 10^2$ s$^{-1}$ | Ref. [45] |
| $w_q = 8\pi C_{Er} N_q N_0$ | $8.1 \times 10^{-19} N_0$ s$^{-1}$ | Ref. [45] |
| $w_{32}$ | $4.2 \times 10^5$ s$^{-1}$ | Ref. [39] |
| $w_{43}$ | $1 \times 10^7$ s$^{-1}$ | Ref. [39] |
| $w_{54}$ | $1 \times 10^7$ s$^{-1}$ | Ref. [39] |
| $C_{b1}$ | $3 \times 10^{-15}$ cm$^3$ s$^{-1}$ | Ref. [39] |
| $C_{up}$ | $7 \times 10^{-17}$ cm$^3$ s$^{-1}$ | Ref. [39] |
| $C_{b2}$ | $3 \times 10^{-19}$ cm$^3$ s$^{-1}$ | Ref. [39] |
| $C_{b3}$ | $3 \times 10^{-19}$ cm$^3$ s$^{-1}$ | Ref. [39] |
| $C_A$ | $3 \times 10^{-19}$ cm$^3$ s$^{-1}$ | Ref. [39] |
| $C_3$ | $7 \times 10^{-17}$ cm$^3$ s$^{-1}$ | Ref. [39] |

In the model we take into account also the possibility that a direct absorption of a 488 nm photon by a Er atom leads to a transition from the ground state to the $^4F_{7/2}$ level. This process is characterized by an excitation cross section $\sigma = 1 \times 10^{-19}$ cm$^2$ (Ref. [39]).

**Deterministic Model for the Si Nanocrystal-Er Atom Interaction**

By considering all the processes represented in the scheme of Figure 1, it is possible to write down a set of rate equations of first order describing the time evolution of the concentration of Si nc's and Er$^{3+}$ ions in each energy level

$$\frac{dn_b}{dt} = \sigma_{ab}\phi n_a - w_b n_b - \sum_{i=1}^{3} C_{bi} n_b N_i \tag{1}$$

$$\frac{dn_a}{dt} = -\sigma_{ab}\phi n_a + w_b n_b + \sum_{i=1}^{3} C_{bi} n_b N_i \tag{2}$$

$$\frac{dN_5}{dt} = \sigma\phi N_1 + \sum_{i=2}^{3} C_{bi} n_b N_i + C_3 N_3^2 - (w_{51} + w_{54})N_5 \tag{3}$$

$$\frac{dN_4}{dt} = C_{b1} n_b N_1 + C_{up} N_2^2 + w_{54} N_5 - w_{43} N_4 \tag{4}$$

$$\frac{dN_3}{dt} = w_{43} N_4 - (w_{32} + w_{31})N_3 - C_{b3} n_b N_3 - 2C_3 N_3^2 \tag{5}$$

$$\frac{dN_2}{dt} = w_{32} N_3 - (w_{21} + w_q)N_2 - C_{up} N_2^2 - C_{b2} n_b N_2 - C_A n_b N_2 \tag{6}$$

$$\frac{dN_1}{dt} = (w_{21} + w_q)N_2 + C_{up} N_2^2 + C_A n_b N_2 + w_{31} N_3 + C_3 N_3^2 - \sigma\phi N_1 - C_{b1} n_b N_1 \tag{7}$$

where $\phi$ is the flux of photons incident onto the sample, $n_a$, $n_b$ and $N_i$, with $i = 1,2,3,4,5$, are the populations of the different levels of Si nanocrystals and Er$^{3+}$ ions, shown in Figure 1. In particular, $n_0 = n_a + n_b$ and $N_0 = \sum_i N_i$ define the total concentrations of excitable Si nc's and Er atoms, respectively. We note that in Eqs. (1)-(7) the interaction between two energy levels is given by the product of the concentrations of centers whose energy values are those of the levels involved. Therefore the interaction probability will be described by rates which are inversely proportional to the square of the mean interaction volumes. This is due to the fact that a short-range dipole-dipole-like interaction, whose strength depends on the inverse sixth power of the mean distance between the two centers, is introduced.

Moreover, we note that in $Er^{3+}$ the non-radiative deexcitation rates between contiguous levels are larger for transitions from the higher energy levels: $w_{54}, w_{43} \gg w_{32} \gg w_{21}$. This implies that in $Er^{3+}$, the higher energy states are much less populated than the others.

## Stochastic Model for the Si Nanocrystal-Er Atom Interaction

Now we extend the model for the Si nc–$Er^{3+}$ ions interaction [39] above detailed by including a stochastic term that mimics an external source of random fluctuations. In particular this contribution is inserted as a multiplicative noise term, $N_2\mu(t)$, in the rate equation for the energy level 2 of $Er^{3+}$ ions. Eq. (6) is therefore replaced by

$$\frac{dN_2}{dt} = w_{32}N_3 - (w_{21} + w_q)N_2 - C_{up}N_2^2 - C_{b2}n_bN_2 - C_A n_b N_2 + N_2\mu(t) \tag{8}$$

Here $\mu(t)$ describes a colored noise and is given by the archetypal exponentially correlated process, known as Ornstein-Uhlenbeck (OU) process [40]

$$\frac{d\mu(t)}{dt} = -\frac{dt}{\tau_C}\mu(t) + \frac{\sqrt{D}}{\tau_C}\xi(t), \tag{9}$$

where $\xi(t)$ represents a Gaussian white noise within the Ito's scheme with zero mean, correlation function $\langle\xi(t)\xi(t')\rangle = \delta(t-t')$, and intensity $D$, while $\tau_C$ is the correlation time of the OU process. The correlation function of the OU process is

$$\langle\mu(t)\mu(t')\rangle = \frac{D}{2}\exp\left(-\frac{|t-t'|}{\tau_C}\right) \tag{10}$$

and gives $D\delta(t-t')$ in the limit $\tau_C \to 0$.

## Stationary State for the Population $N_2$ and Decay Time

The lifetime of the metastable level 2 has been investigated as a function of both the noise intensity and the correlation time, integrating Eqs. (1)-(5), (7), (8), and averaging over a certain number of numerical realizations.

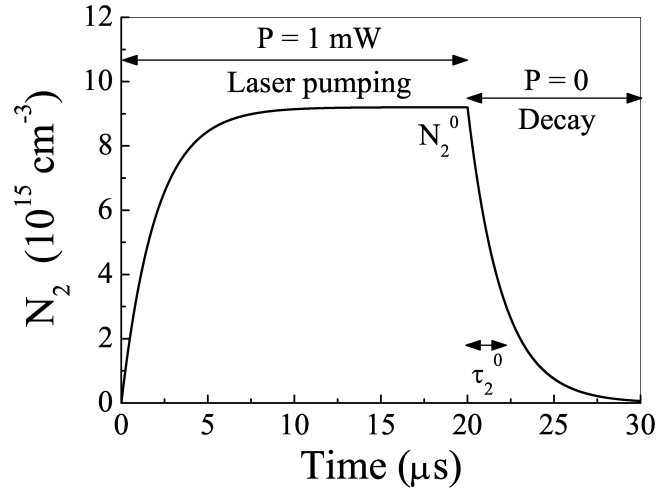

Figure 2. Time evolution of the population of the first excited state of $Er^{3+}$ ions, $N_2$, as a function of time, in the presence of *laser pumping* of power $P = 1$ mW ([0,20] μs) and during the time *decay*, $P = 0$, ([20,30] μs). The laser pump has been switched off after the population $N_2$ has reached the stationary value $N_2^0$. $\tau_2^0$ is the lifetime of the population of $Er^{3+}$ ions in the level 2, obtained in the absence of noise.

In Figure 2 we show the population of Er atoms in the energy level 2, i.e. $N_2$, as a function of time, obtained in the absence of external noise ($D = 0$). In particular, we divide the simulation time in two time windows. In the first one, the numerical integration of Eqs. (1)-(5), (7), (8) is performed using a value of the *laser pumping* power, $P$, equal to 1 mW. In the second time interval, after the system has reached the steady state condition i.e. $N_2 = N_2^0$, the laser beam is switched off ($P = 0$). This determines a time decay of the population $N_2$, and the lifetime $\tau_2$ of the population of $Er^{3+}$ ions in the level 2 is calculated as the time for which a reduction of $N_2^0$ by a factor $e^{-1}$ is observed. For the value of pump power used here (1 mW), this lifetime is approximately equal to *2 ms*. This value obtained in the absence of external noise, will be denoted, from now on, by the symbol $\tau_2^0$. For the case shown in Figure 2, $N_2^0 \cong 9.20 \times 10^{15}$ cm$^{-3}$.

## EFFECTS OF THE EXTERNAL NOISE SOURCE

In Figure 3 we show the population of $Er^{3+}$ ions, $N_2$, in the energy level of the first excited state $^4I_{13/2}$, as a function of the time. The analysis is carried out in the absence of noise ($D = 0$), and for different values of the noise intensity $D$ (panel (a): $D^{1/2} = 10$ s$^{-1}$; panel (b): $D^{1/2} = 10^2$ s$^{-1}$) and noise correlation time $\tau_C$ (= $0.1\tau_2^0$, $\tau_2^0$, $10\tau_2^0$).

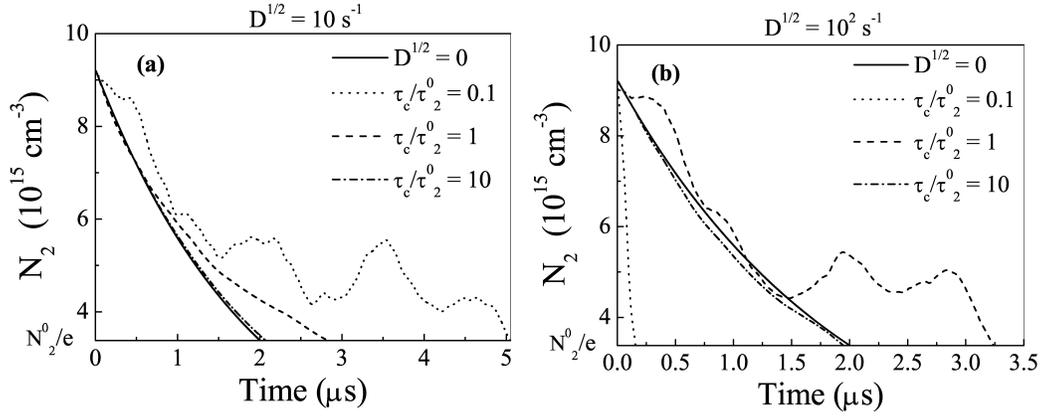

Figure 3. Time evolution of the population of the first excited state of $Er^{3+}$ ions, $N_2^0$, as a function of the time, obtained after the pumping with a laser beam of power 1 mW has been turned off. The curves have been obtained in the absence of noise ($D=0$), and for $D^{1/2}=10$ s$^{-1}$ (panel a), $D^{1/2}=10^2$ s$^{-1}$ (panel b), using three different values of the noise correlation time ($\tau_C = 0.1\tau_2^0, \tau_2^0, 10\tau_2^0$).

For $\tau_C \ll \tau_2^0$ the deexcitation process of $^4I_{13/2}$ level is considerably affected by the external fluctuations, showing different trends which depend on the value of the noise intensity. Specifically, for $D^{1/2}=10$ s$^{-1}$ the external noise causes a slowing down of the decay from the first excited level of Er atoms (dotted-line curve in panel (a) of Figure 3), while determines a faster depletion for $D^{1/2}=10^2$ s$^{-1}$ (see dotted-line curve in panel (b) of Figure 3). This effect is quite counterintuitive since the colored noise has a negligible memory (small value of $\tau_C$) with respect to the characteristic decay time $\tau_2^0$ of the first excited level of Er atoms. A different behaviour should be observed if $Er^{3+}$ ions were isolated. Conversely, in our system the Er atoms interact with the Si nc, exchanging energy, and the transition $^4I_{13/2} \rightarrow \,^4I_{15/2}$ competes with the deexcitation processes from higher energy levels, e.g. $^4I_{11/2} \rightarrow \,^4I_{13/2}$ characterized by a rate $w_{32} = 4.2 \times 10^5$ s$^{-1}$. Concerning this, one can note that the inverse of the noise correlation time, $\tau_C = 0.1\tau_2^0$, is of the same order of magnitude of the transition rate $w_b$ between the energy levels $b$ and $a$ of the Si nc. These are indirectly affected by fluctuations through the energy exchanges with the Er levels. As a consequence, for short memory values, the time evolution of the energy level 2 of $Er^{3+}$ ions is influenced in a qualitatively different way by the external noise source, and dependent on the values of $D$.

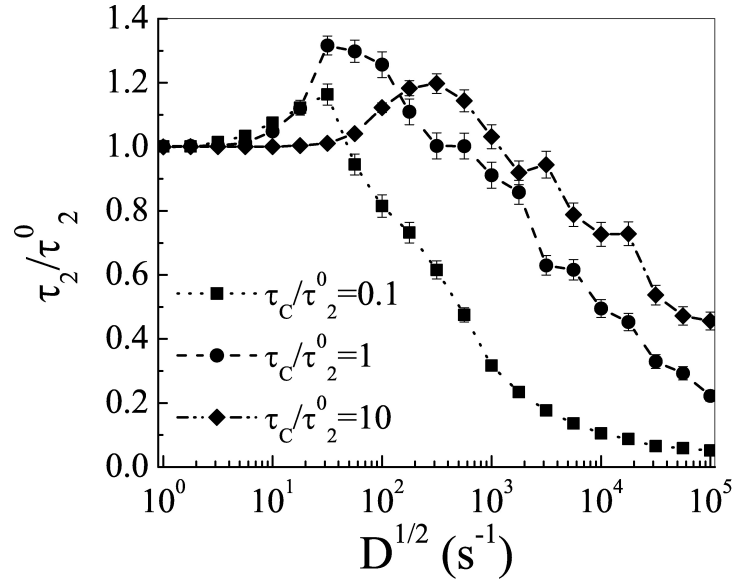

Figure 4. Ratio between the lifetime of the first excited level of the $Er^{3+}$ ions $\tau_2$ in the presence of noise and $\tau_2^0$ (obtained in the absence of noise) as a function of the noise amplitude of the external colored noise $D^{1/2}$. Curves were obtained for different values of $\tau_C$, namely, $\tau_C = 0.1\tau_2^0$, $\tau_2^0$, $10\tau_2^0$.

For $\tau_C \sim \tau_2^0$, both values of the noise intensity, $D^{1/2} = 10$ s$^{-1}$ and $D^{1/2} = 10^2$ s$^{-1}$, determine a clear slowing down of the deexcitation process of the level 2 of Er atoms (see dashed-line curves in both panels of Figure 3). Differently, when $\tau_C \gg \tau_2^0$, the deexcitation process becomes quasi-deterministic and the curves (dashed-dotted line in both panels of Figure 3) of the population $N_2$ approach those obtained in the absence of external noise ($D^{1/2} = 0$).

To understand better the effects of the external colored noise on the deexcitation process of the metastable level $^4I_{13/2}$, we evaluated (performing $10^4$ numerical realizations) both the average values and error bars of the deexcitation times of the $^4I_{13/2}$ energy level.

In particular, for the $^4I_{13/2}$ level of the Er atoms, we calculate the ratio between the deexcitation time $\tau_2$ in the presence of noise and that obtained in the absence of noise ($\tau_2^0$), as a function of the amplitude of the external colored noise, $D^{1/2}$, and for different values of the ratio $\tau_C/\tau_2^0$, namely $\tau_C/\tau_2^0 = 0.1, 1, 10$. The results, shown in Figure 4, indicate that the addition of a source of correlated noise, characterized by $0.1\tau_2^0 < \tau_C < 10\tau_2^0$, can determine an enhancement of the lifetime $\tau_2$ with respect to $\tau_2^0$ up to 30% for $\tau_C = \tau_2^0$. We note that $\tau_2/\tau_2^0$ is a nonmonotonic function of $D^{1/2}$, which exhibits a maximum whose position depends on the value of the correlation time $\tau_C$. In particular the position of the maximum is shifted towards right for increasing values of $\tau_C/\tau_2^0$. The nonmonotonic behaviour of $\tau_2/\tau_2^0$ as a function of $D^{1/2}$, which represents a further example of noise enhanced stability (NES) [17-21, 32-38], determines an improvement of the upconversion process. In order to quantify the effect of the external noise on the upconversion mechanism, we introduce the quantities $\eta$ and $\eta_0$ defined as the efficiency of the upconverting process from the level $^4I_{13/2}$ in the presence and in the absence of noise, respectively. The effect of noise on the upconversion mechanism can be therefore evaluated defining the efficacy $\eta/\eta_0$. Pilla et al. and Jacinto et al. have showed that the quantum efficiency of the fluorescence of a metastable state is related to the speed of ETU processes. [9-10] In particular they have obtained the following relation between the quantum efficiency and the lifetime of a metastable state

$$\eta = \frac{w_{21}\tau_2}{1+C_{up}N_2^0\tau_2}, \tag{11}$$

valid also in the absence of noise with the foresight to substitute $\tau_2$ with $\tau_2^0$.

Using Eq. (11) to express the ratio $\eta/\eta_0$, one gets

$$\frac{\eta}{\eta_0} = \frac{\tau_2}{\tau_2^0} \cdot \frac{1+C_{up}N_2^0\tau_2^0}{1+C_{up}N_2^0\tau_2}. \tag{12}$$

Since the binomial terms are practically equal to 1, the ratio $\eta/\eta_0$ coincides with $\tau_2/\tau_2^0$, so that the latter provides a good measure of the efficacy for the system investigated. As a consequence, the results shown in Figure 4 represents the behaviour of the efficacy $\eta/\eta_0$ as a function of the noise amplitude $D^{1/2}$, for different values of the noise correlation time.

## CONCLUSION

In this chapter we have investigated the influence of an external source of colored noise on the deexcitation process of the energy level $^4I_{13/2}$ in Er-doped Si nanocrystals, a system which is strongly interested by the upconversion phenomenon. Our findings show that the inclusion of a source of multiplicative colored noise, which acts on the population of the level $^4I_{13/2}$ of the Er atoms, can slow down the deexcitation process. As a consequence, the lifetime may increase up to 30% of the value observed in the absence of external noise. This enhancement, which is nonmonotonic with the amplitude of the external fluctuations, presents a maximum within a wide interval of intensities and correlation times of the noise source. The slowing down of the deexcitation process can be ascribed to the effective deexcitation rates of the Er atoms, during the depletion dynamics of $^4I_{13/2}$, i.e. the first excited level of $Er^{3+}$ ions.

In conclusion, the enhancement of the deexcitation time of the level $^4I_{13/2}$ of $Er^{3+}$ ions embedded in Si nanocrystals, is strongly connected with both the correlation time and intensity of the noise source. Therefore, external fluctuations play a key role in modulating the average lifetime of the metastable state $^4I_{13/2}$, and can contribute to increase the efficiency of upconversion phenomena, such as the cooperative upconversion between Er atoms. By adjusting the intensity of the external noise, it is possible to select the most favorable working condition in view of obtaining more efficient upconverters integrated in photovoltaic devices.

## ACKNOWLEDGMENTS

This work was supported by MIUR through Grant. No. PON02_00355_3391233, "Tecnologie per l'ENERGia e l'Efficienza energeTIC - ENERGETIC"..